# Ameliorating transport system focusing on sustainability and inclusiveness through a mixed-method research (A case study in Tehran, Iran)

Melika Soufiemami [1]

Department of urban planning, Shahid Beheshti University, Tehran, Iran

## 1. Introduction

Sustainable transportation is an aspect of sustainable development that in case of its establishment, several urban issues will be improved or even eliminated. Sustainable transport planning (STP) is a mean to decrease noise and air pollution, car accidents, traffic congestion, carbon footprint, the cost of inner city travel, and many other problems. Also it can contribute to ameliorate the condition of accessibility in urban areas, justice between different groups of dwellers, safety and etc., [1,2,3].

In some parts of the world, motorized transport has undermined human health and has been a burden to sustainability [4,5], while some cities are developing more active modes by following paradigms like "growth to equity and sustainability" [6]. Exploiting urban infrastructure is a right for all citizens regardless of their gender, wealth, age, health condition and etc., which is called "the right to the city" [7] and obviously the right to accessibility seems to be of an indispensable importance for all urban dwellers.

---

[1] Email address: Melika.soufi@gmail.com
Address: No. 412, Block A, Hayat-e-sabz building, 4th Tangestan alley, Pasdaran street, Tehran, Iran.

Challenges for excluded people in terms of public transport (PT) have been mitigated over time [8] and numerous experts focused on gender needs in STP which is undoubtedly an important topic, however, we know little about the needs and hindrances faced by disadvantaged groups of people are encountering every day.

In this paper, I have adopted four groups as disadvantaged people to assess their transportation needs in the context of Tehran city, the capital city of Iran. These groups include disabled individuals (usually suffering from mental or physical incapability) [9], children who are not aware of their urban rights and might not be able to express their needs, elderlies (people aging more than 65) and women as citizens with socio-cultural burdens, especially in an Islamic country like Iran. This process would contribute to ameliorate sustainable and inclusive transport planning in addition to providing a greater quality of life for all people.

## 2. Literature review

A variety of articles on the current literature have focused on policy, governance, or design for one of the excluded groups such as elderlies or disabled people. [10] catechized transport and disability in order to eliminate social exclusion. Many articles have investigated on gender-sensitive inclusion in STP [11,12,13]. generally, considering all of the disadvantaged groups (children, women, elderlies and disabled people) is known as a gap in inclusive transport studies and focusing on them must be undertaken.

### 2.1. Sustainability and inclusiveness

there is not a clear and simplified definition of inclusive planning [14], however, STP emphasizes on three main features: economic, social and environmental aspects [15]. In other words, it allows justice between generations without any dispute [16], reduces greenhouse gasses, air and noise pollution, poverty and preserves economic growth [17,18,19]. The social aspect of STP however insists on an inclusive transportation services access for all [20,18]. Inclusive transportation focuses on vulnerable and disadvantaged urban dwellers with the purpose of life standard achievements [21]. For example, 1.4% of Iranians are disabled dwellers [22], who must be given the right of initial public service [23] and if ill-equipped decision continuous in this societies, there soon will be social heterogeneity, and many other challenges [24] the result of ignorance of the inclusiveness might be felt among inhabitants who are feeling lack of land service, insecurity and tension, deregulation of public space and traffic congestion [25].

### 2.1.1. disabled people

Tehran, the capital of Iran, which is known as a megacity with a population of 8.693.706 has been made for motorized vehicles, while a city must have standards for the needs of all groups [26] ranging from cyclist, elderlies, disabled individuals, students, people in poor economic condition. Disadvantaged people are tackling various problems ranging from lack of accessibility to insecurity or safety obstacles. Unfortunately, they are being ignored in many cases. For instance, the number of disabled people had not been counted in the latest notional census of Iran in 2016.

### 2.1.2. women

Women in Iran, face certain limitations about the obligatory dressing code known as "Hijab" which does not allow them to enjoy some simple daily physical activities such as cycling [27,28,29,30]. Thus, focusing on minorities such as: women, children, elderlies and disabled people [31,32,33,34] must not be ignored. Another cross-cutting issue for women is the fear of violence and crime in urban spaces according to some dangers presented to their life and their mobility might unwittingly reduce [35,36,37,38,39,40,41].

### 2.1.3. Elderlies

people over 60 will be around twelve million by 2030 [42] and there seems to be a growing correlation between age and disabilities [43]. Thus, the need for an inclusive planning which emphasizes on population aging is necessary [44]. Elderly people usually have a fear of falling [45], need for bathrooms during daily commutes "slow modes of transport such as cycling" [46], and mobility which is more important than accessibility because it can increase the level of walkability and physical activity [47,48].

### 2.1.4. Children

A reduction physical activity among children is occurring [49,50], so there must be solutions such as active modes of transport to this dramatic concern. However, accessibility and mobility ought to have certain features for children, including; safety, security, street connectivity, special facilities for walking and cycling.

## 3. case study and conceptual model

Municipality of Tehran has divided the city into 22 administrative districts. however, in this work, I have examined district number 8 which is presented in Fig 1. .

Tehran's (PT)includes variety of PT modes such as bus, BRT, subway. Albeit, private modes of transport (taxi, fix-route taxi, hailing taxies) have grown their popularity among Tehran dwellers [52]. A conceptual model was according to literature review and the context of the case study shown in Fig 2. .

## 4. Data collection and methodology

The chosen method adopted for this study is a mixed research method in which qualitative and quantitative methods are used to contribute the robustness of the results. According to the available data, literature review and context of the case study, 6 factors were chosen (Low carbon transport, Acceptable accessibility, Cost-effectiveness and affordability, Promoting cycling and walkability, Usability of PT for Disadvantaged people, PT diversity). Each factor contains one or some indicators and colleting method which are presented in table 1. These indicators were assessed through 180 face to face questionnaire, 9 focus groups including disadvantaged dwellers, key informants' interviews and map analysis by means of ArcGIS, and finally available data in Portal of Tehran municipality (https://Tehran.ir), and comprehensive plan of Tehran.

## 5. Analysis and results

Based on table 1, the order of analysis will be revealed in this section of the study.

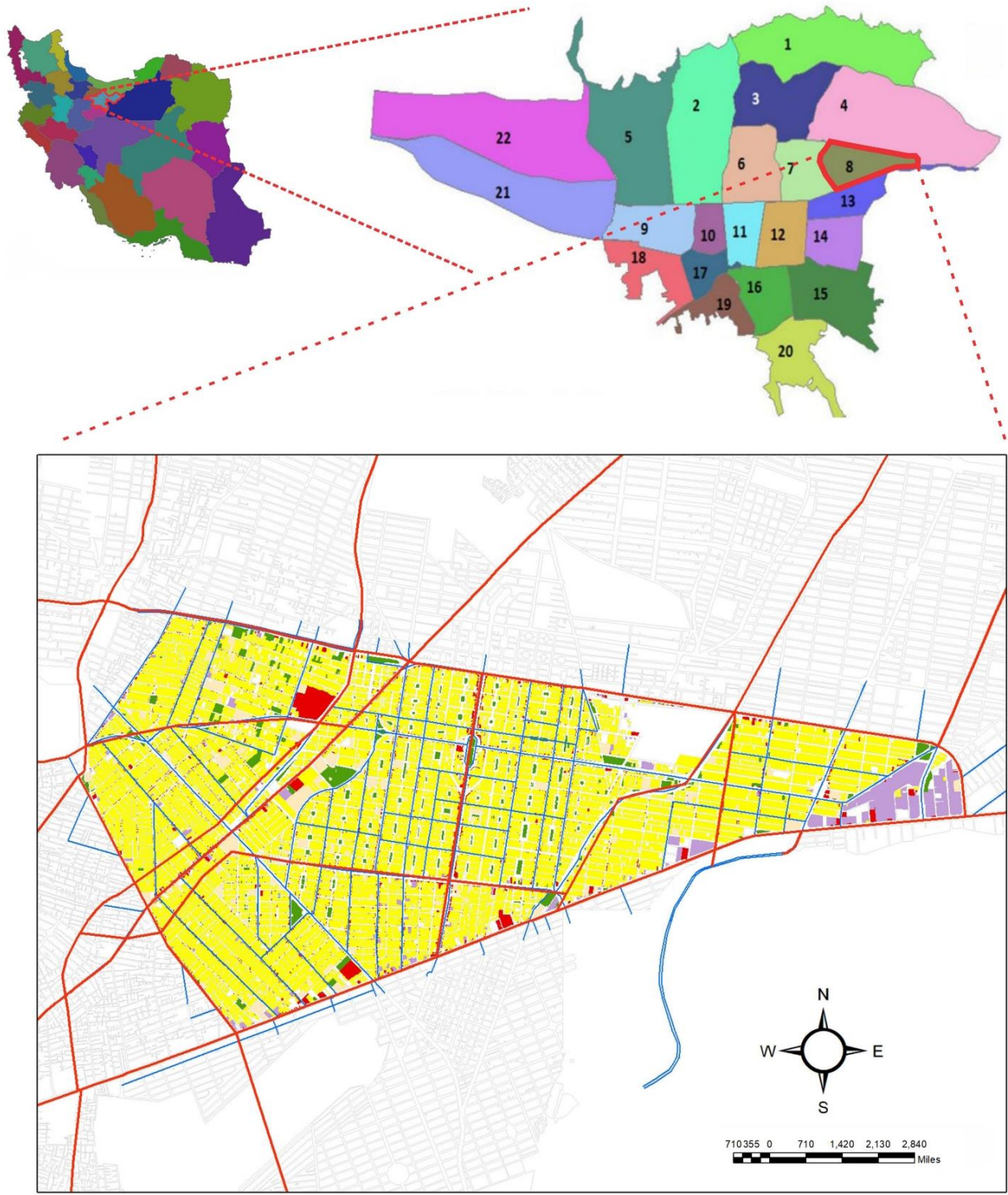

*Fig. 1. Two maps on the top show Iran and Tehran city from left to right, and the map below shows district 8 of Tehran.*

### 5.1. Low carbon transport

Fossil fuels consumption has a correlation with carbon emission and air pollution. The comparing level of AQI of Tehran city and district 8 as a factor related to low carbon transport system is demonstrated in table 2. It is clear that the case study area is less polluted than other districts on average.

## 5.2. Acceptable accessibility

Three indicators including number of transportation stations in the area and the approximate distance between stations and basic urban service land uses were chosen for this factor. Basic urban land use for disadvantaged people are: primary schools, shopping centers, parks and public transportation stations. Disadvantaged people are capable to walk for 10 minutes without being exhausted. In this part of study, I have analyzed the distance between mentioned land use in the case study area by the means of network analysis[2] in ArcGIS which are shown in Fig. 3 to Fig.6.

The accessibility for educational centers seems to be ideal for the central part of the district, while eastern and some western part of the district lack in accessibility to these centers. While, in case of parks, shopping centers and public transportation platforms, the network system covers roughly all of the case study area.

## 5.3. Cost-effectiveness and affordability

Two questions were prepared in a 3-point Likert scale to estimate cost-effectiveness from the aspect of time-saving, and financial affordability of monthly PT costs. 180 people participated in the questionnaire and the results indicate that; the PTS is roughly wastes dwellers time, because more than half of the participants were not eager to use public transportation because of shortage in time-cost-effectiveness. Conversely, most of them were satisfied with the monthly cost which they pay for public transportation fees. The results of the questionnaire are available in Fig.7.

## 5.4. Promoting cycling and walkability

Four of the indicators are assessed through the focus groups. Number of 9 focus groups consist of elderlies who were suffering from one or more than a disability, women and children were made. People pointed out problems in terms of cycling and walkability, and the interview was extended around the using a bicycle or walking for short daily trips. According to an interview with the mayor there is not any bicycle lane in the area. Results of the focus groups are available in table 3.

## 5.5. Usability of PT for Disadvantaged people

Number of subway stations with elevator and textile paving for elderly and disabled groups is illustrated in Fig. 8. According to this map, only 3 stations are usable for people with blindness and 5 stations are proper for disability use. The result of the focus groups for hindrances faced by disadvantaged groups are lack of elevator for level change in streets, pedestrian bridges, stations and etc., certain limitations of cycling for women, lack of bicycle lanes in the area, safety issues and fear of accident with motorized vehicle while passing the street during walking or cycling and level change on streets and stations.

## 5.6. PT diversity

There is a variety of public and private modes of transportation is available in this district, including 8 subway stations, 23 BRT stations and more than 29 bus stops. Variety of public transportation platforms in the case study area are depicted in Fig.9.

---

[2] Network analysis (Finding Coverage), n this type of network analysis, drive-time areas correspond to the distance that can be reached within a specific amount of time.

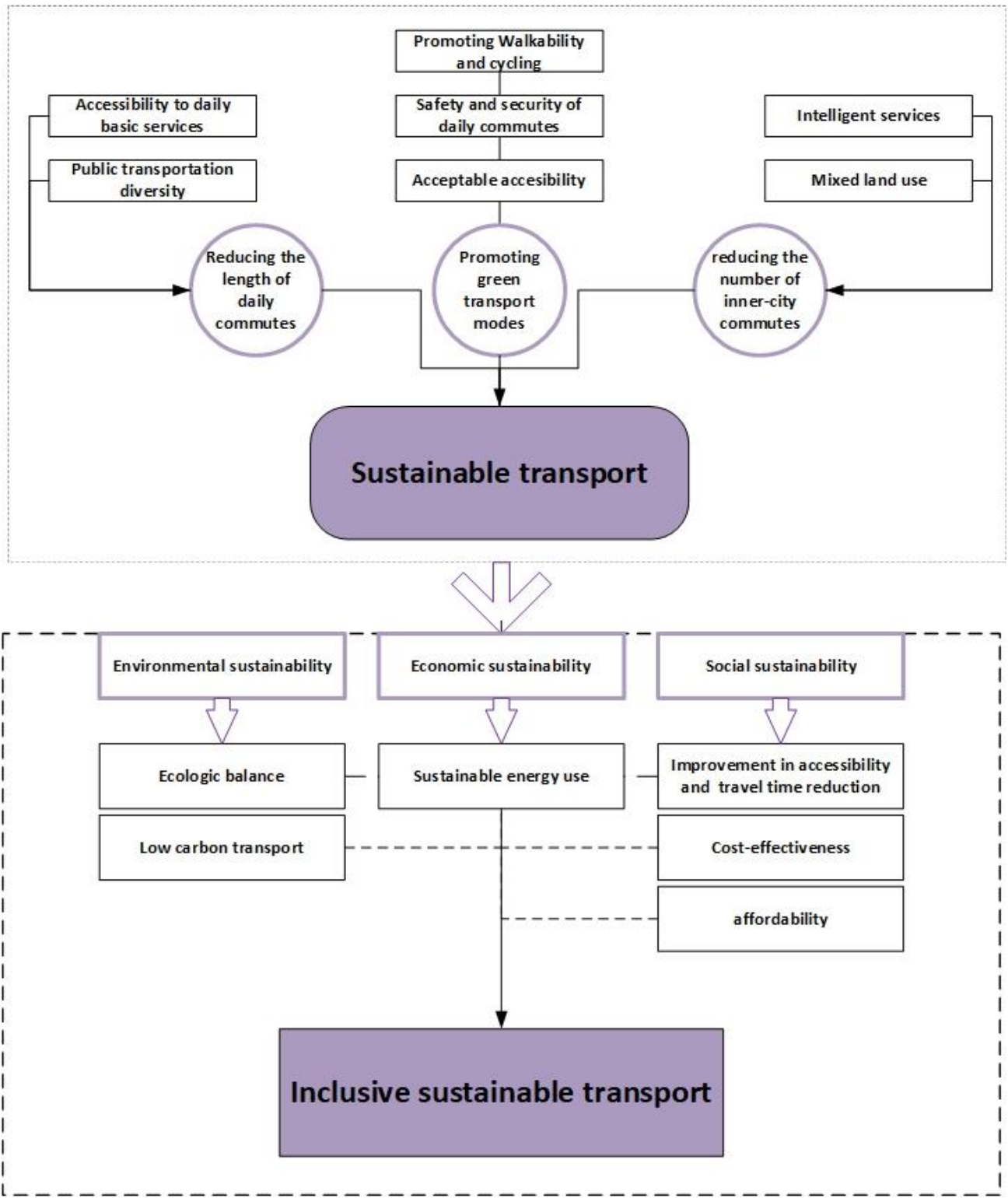

Fig. 2. the conceptual model of the study.

Table 1. Factors, indicators and the chosen assessing method.

| No. | Factor | Indicator(s) | method |
|---|---|---|---|
| 1 | Low carbon transport | Air quality index (AQI) in the study area within a year compared to Tehran city<br>The amount of fossil fuel consumed by private mode of transport | Available data<br>Key informant interview<br>Available data |
| 2 | Acceptable accessibility | The distance between basic urban service and public transportation platforms | Map analysis |
| 3 | Cost-effectiveness and affordability | Average time spent on waiting for a (PT)vehicle arrival<br>Monthly amount of money spent on commuting by public transportation | Questionnaire<br>Questionnaire |
| 4 | Promoting cycling and walkability | Security and safety along bicycle lanes and pavements especially for disadvantaged people<br>Tendency and possibility of cycling and walking among disadvantaged people | Focus group<br>Key informant interview |
| 5 | Usability of (PT) for Disadvantaged people | Number of subway stations suitable for use of disable groups<br>Hindrance faced by people with wheelchair, luggage, etc. to public transportation platforms<br>Hindrance faced by people in level change such as stairs, overpass, etc.<br>Safety and security of disadvantaged groups during commute | Map analysis<br>Focus group<br>Focus group<br>Focus group |
| 6 | (PT)diversity | Availability of bus, BRT and subway within a walking distance | Map analysis |

Table 2. comparing AQI in Tehran city and district 8 in 4 consecutive years.

| year | 2019 | 2020 | 2021 | 2022 | Mean AQI of four years | Description of AQI |
|---|---|---|---|---|---|---|
| AQI of Tehran city | 84 | 128 | 112 | 119 | 110.75 | Unhealthy for sensitive groups |
| AQI of district 8 of Tehran | 58 | 125 | 69 | 97 | 87.25 | Moderate |

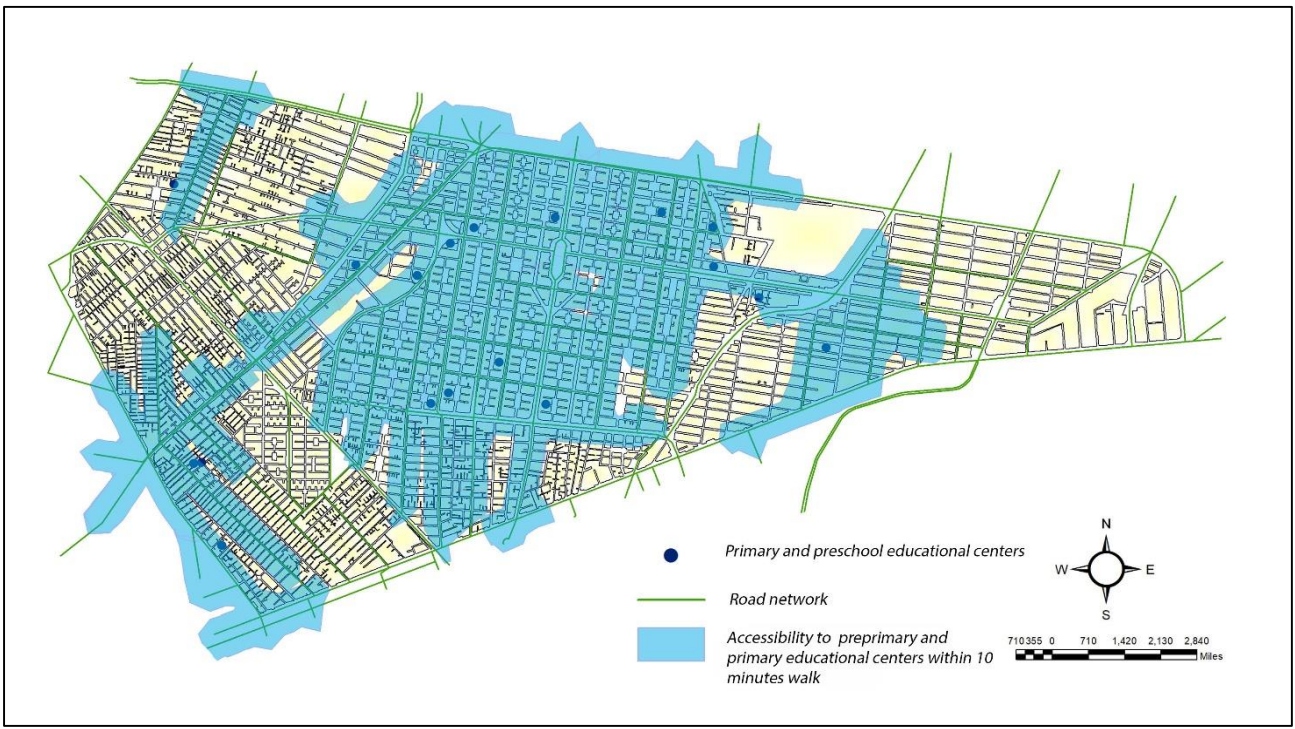

Fig. 3. The accessibility and mobility to primary schools in the area within 10 minutes' walk. .

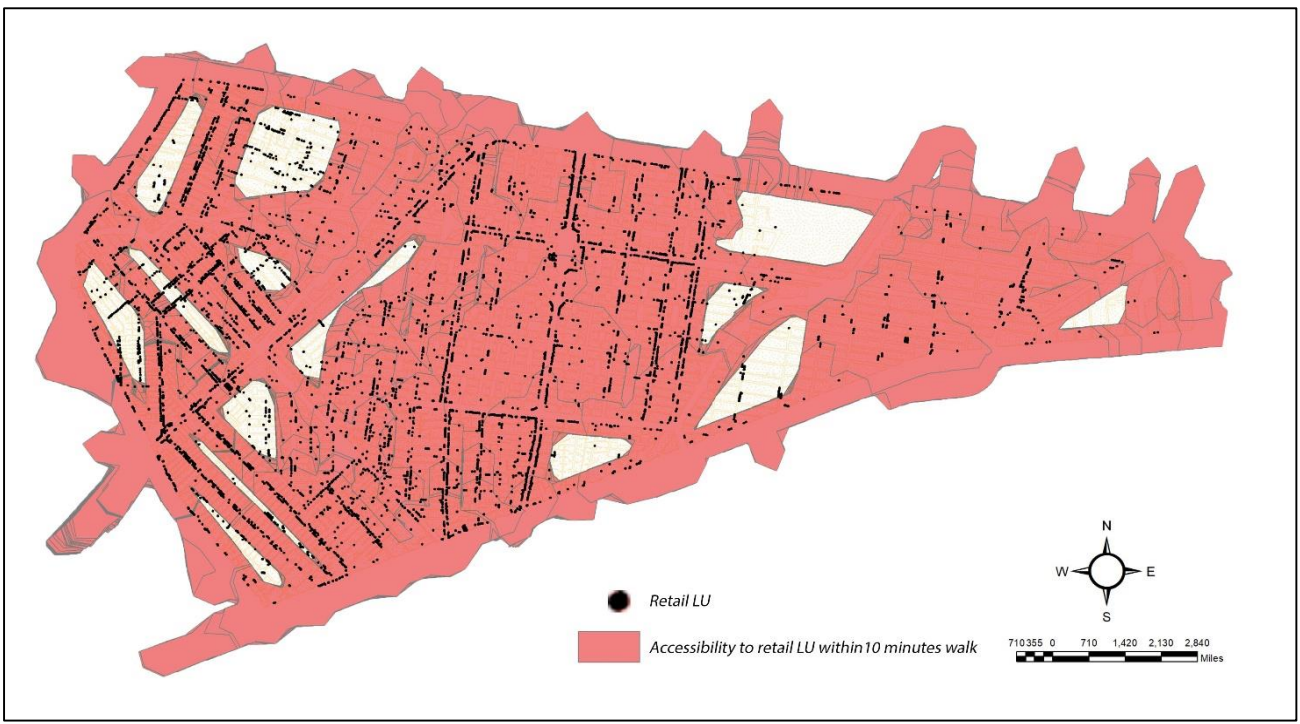

Fig. 4. The accessibility and mobility to shopping centers in the area within 10 minutes' walk.

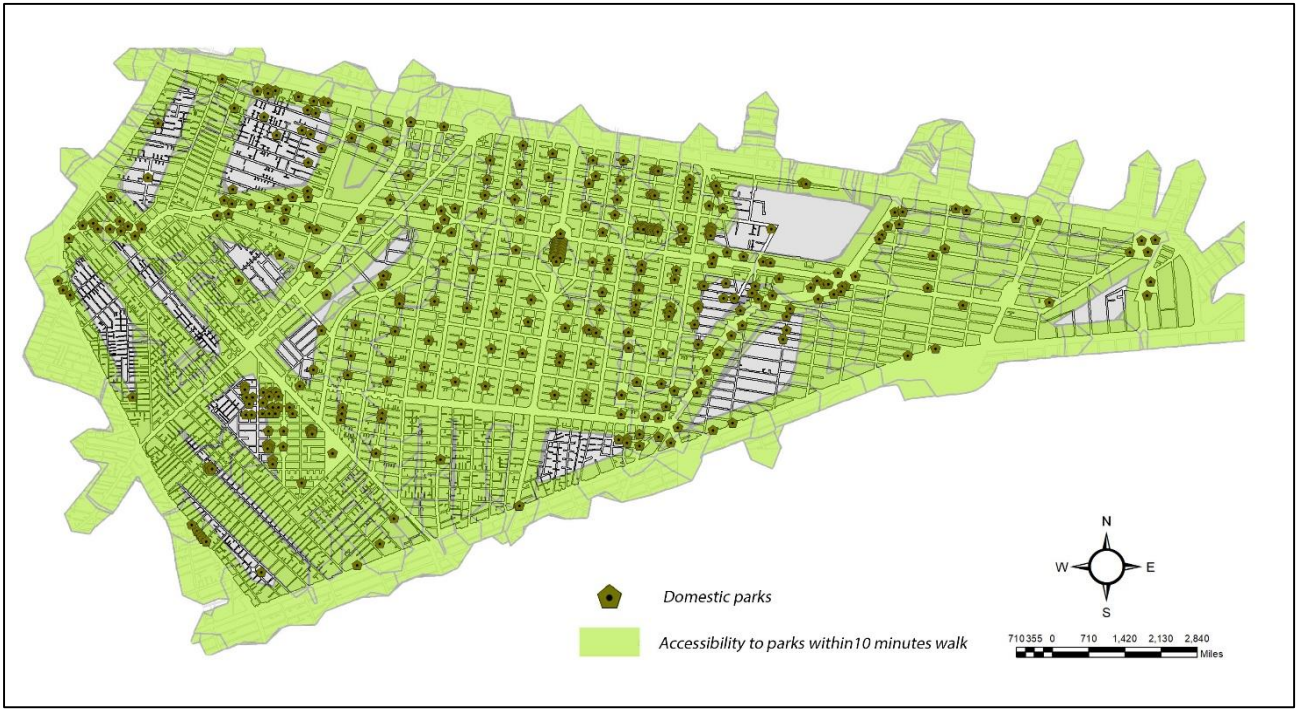

*Fig. 5. The accessibility and mobility to parks in the area within 10 minutes' walk.*

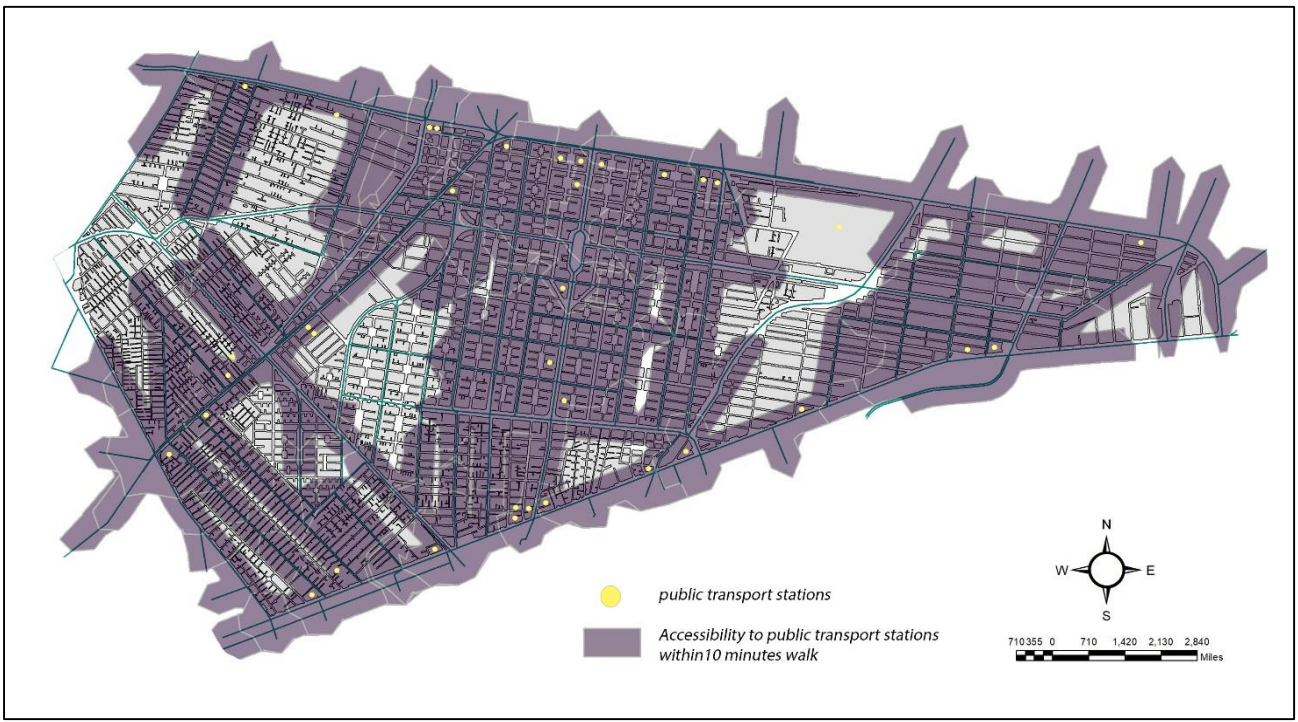

*Fig. 6. The accessibility to public transport stations in the area within 10 minutes' walk.*

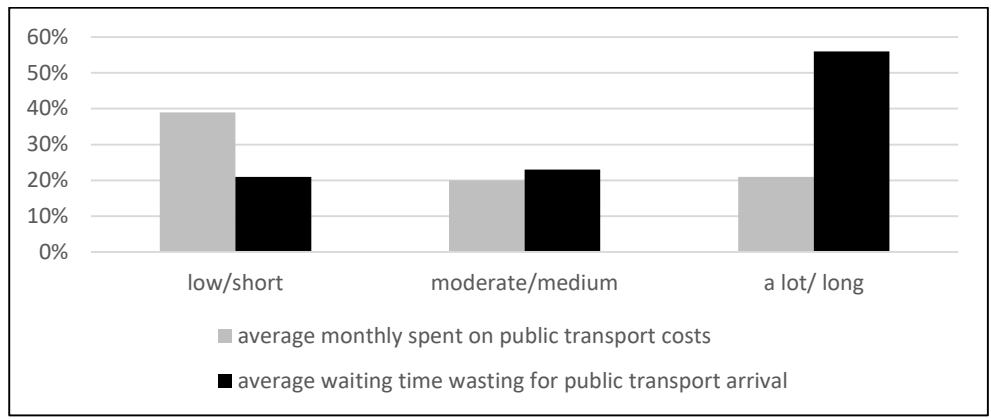

*Fig. 7. results of the questionnaire.*

*Table 3. Results of the focus groups.*

| factor | Issue/outcome | Number of times/issues were identified |
|---|---|---|
| **Promoting cycling and walkability** | Fear of accident in disadvantaged people while using a bicycle | 32 |
| | Lack of bicycle lanes | 40 |
| | Banning women from using bicycles | 45 |
| **Usability of PT for DGs** | Absence of a ramp for getting in to buses or BRTs for disabled and elderly people | 23 |
| | The lack of elevators and routes for the blind in some subway stations | 46 |

*Table 4. Problems revealing in the study and given recommendations.*

| Problems | recommendations |
|---|---|
| Accessibility problems in some parts of the case study | Provision of land use per capita |
| Cost-ineffectiveness of the PT system | Location allocation of PT stations in needed neighborhoods of the area |
| Inexistence of bicycle lanes | Design a complete street in a suitable location, and bicycle lane |
| Safety problems and fear of car accidents | Reducing the width of the street and increasing sidewalks width |
| Apathy of active modes of travel use | Location allocation of a park and ride system |

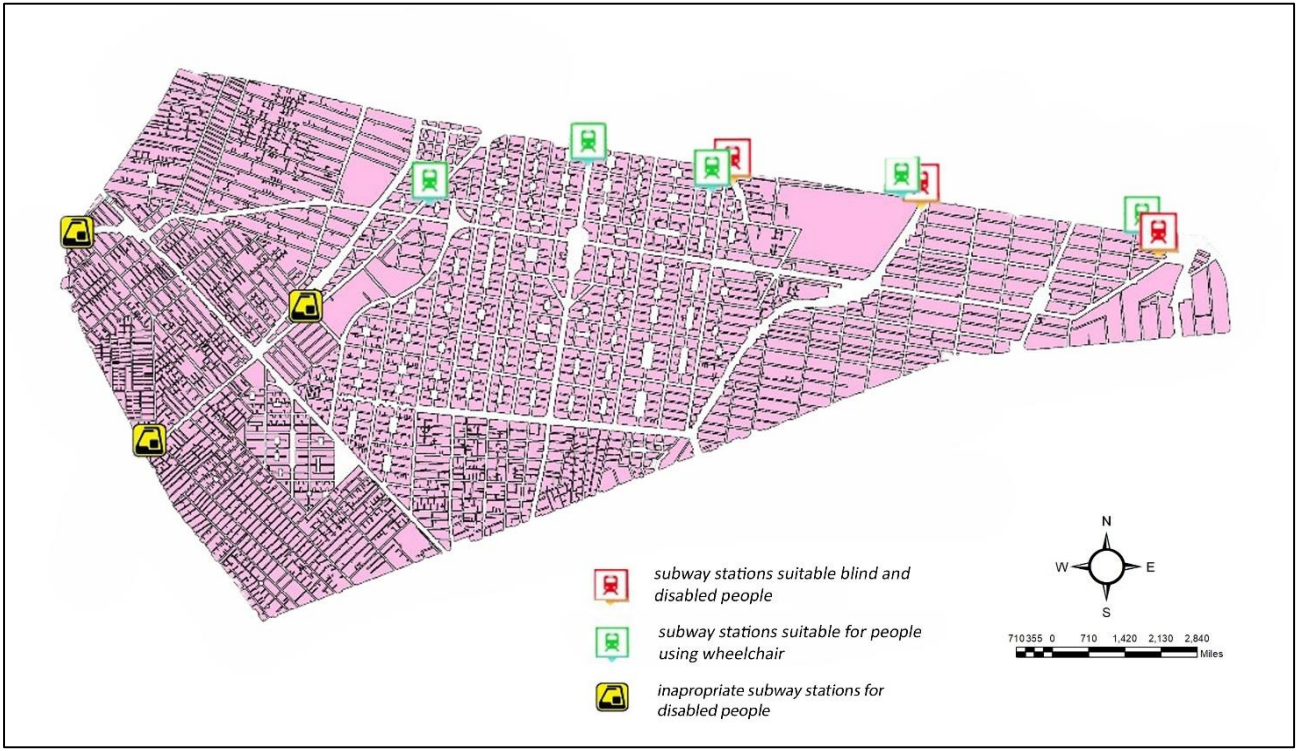

*Fig. 8. Subway stations equipped for disabled and blind individuals, and inappropriate subway stations for them.*

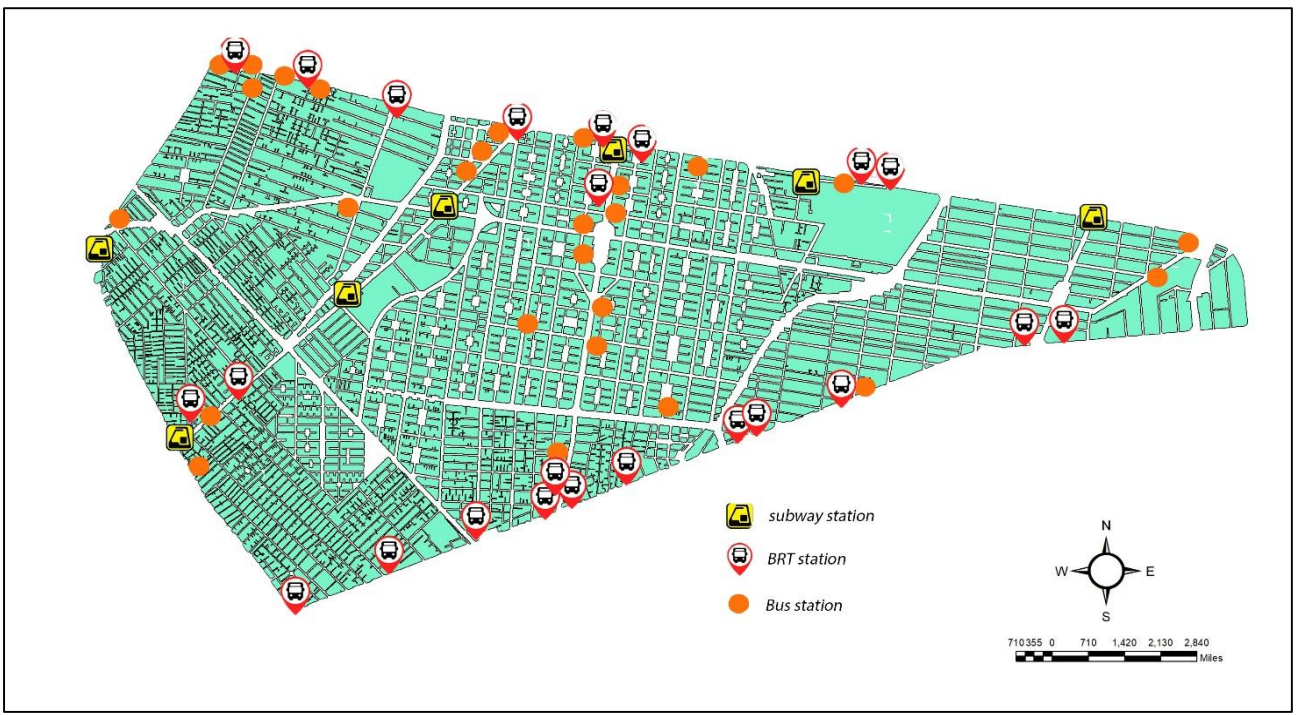

*Fig. 9. Variety of PT stations including subway, BRT and bus stations.*

## 6. Recommendations

problems in the study area vary in terms of sustainable and inclusive transport. Based on the achieved results in section 5, a number of recommendations ranging from provision of land use per capita (for instance for educational centers such as primary schools), Location allocation of PT stations for neighborhoods which have transit deserts [53], designing a complete street and special active modes of transport lanes in a suitable location, promoting active modes such as walking by increasing sidewalks width and Location allocation of a park and ride system.

## 7. Conclusions

In this paper sustainability and inclusiveness of transportation system in district 8 in Tehran was explored. A mixed method research including a face to face questionnaire, 9 in person focus groups, key informant interview, available data analysis and map analysis was applied.

The results indicate that the transport system is low carbon in the case study area comparing to Tehran city. However, it does not have the maximum standard of the air quality yet. In terms of inclusiveness of transport system, there was found that disadvantaged people face hindrances while using public transport modes. These burdens range from lack of elevators and ramps while level change, lack of textile paving for blind individuals, fear of falling down and car accidents while walking and apathy to use active modes of transport because safety problems and specific social limitations for women for cycling. In addition, dwellers in this district, were not satisfied with cost-effectiveness of PT. conversely they were pleased with the affordability of PT. The results also show that active transport modes such as cycling and walking are not popular among citizens due to lack of infrastructures such as specific bicycle lanes. Thus, the transport system in the study area needs to ameliorate which can be done through various actions such as providing basic land use per capita, allocating PT stations for transit deserts, location allocation and designing a park and ride system.

To conduct this study in another context, it should be noticed that; although needs of excluded people might be approximately resembling, sustainability of the transport system might differ from city to city. Hence, this study may not be directly generalizing in another context, however, it has valuable results which can ameliorate quality of life for all people, enhance air quality, social equity and finally lead to a low carbon city.

## 8. Declaration of competing interest

The author declare that she has no competing financial interests or personal relationships that could have appeared to influence the work reported in this paper.